\documentclass[aps,showpacs,preprint,footinbib,preprintnumbers]{revtex4}

\usepackage{graphicx}
\usepackage{dcolumn}
\usepackage{bm}
\usepackage{longtable}

\begin{document}


\title{Heavy quark dynamics for charm and bottom flavor \\ on Fermi surface at zero temperature}


\author{S.~Yasui}
\email[]{yasuis@post.kek.jp}
\affiliation{KEK Theory Center, Institute of Particle and Nuclear
Studies, High Energy Accelerator Research Organization, 1-1, Oho,
Ibaraki, 305-0801, Japan}
\author{K.~Sudoh}
\affiliation{Nishogakusha University, 6-16, Sanbancho, Chiyoda,
Tokyo, 102-8336, Japan}


\date{\today}

\begin{abstract}
We discuss the dynamics in finite density medium including a heavy impurity particle (hadron or quark) with a heavy flavor, charm and bottom, at zero temperature.
As a system, we consider a $\bar{D}$ ($B$) meson embedded in nuclear matter as a heavy impurity boson with SU(2) isospin symmetry.
As another system, we also consider a charm (bottom) quark embedded in quark matter as a heavy impurity fermion with SU(3) color symmetry.
We suppose a vector current interaction with SU($n$) symmetry ($n \ge 2$) for the fermion composing the Fermi surface and the embedded heavy impurity particle, and calculate the scattering amplitude perturbatively for the small coupling constant up to one-loop level.
We obtain that the scattering amplitude has a logarithmic enhancement in the large mass limit of the heavy impurity particle, and show that the perturbative calculation breaks down for any small coupling constant in this limit.
\end{abstract}


\maketitle

\section{Introduction}

In the present hadron and quark physics, heavy quarks are interesting subjects for understanding the properties of QCD.
In recent literature, there have been many discussions about charmed and bottom nuclei with bound charm and bottom hadrons.
For example, charmed nuclei may have $\Lambda_{\rm c}$ baryon \cite{Tsushima:2002ua,Tsushima:2002sm,Tsushima:2003dd}, $\Sigma_{\rm c}$ baryon \cite{Tsushima:2003dd}, $D$ meson \cite{Mizutani:2006vq,Bayar:2012dd,Oset:2012px},  $\bar{D}$ meson \cite{Tsushima:1998ru,Sibirtsev:1999js,Tsushima:2002cc,Hilger:2008jg,Hilger:2010zb,Wang:2011mj,Mishra:2003se,Lutz:2005vx,Tolos:2007vh,Mishra:2008cd,Kumar:2010gb,JimenezTejero:2011fc,Kumar:2011ff,GarciaRecio:2011xt,Yasui:2012rw} for open charm and $J/\psi$  \cite{Klingl:1998sr,Song:2008bd,Morita:2010pd,Hayashigaki:2000es,Friman:2002fs} for hidden charm, inside normal nuclei.
Bottom nuclei may have $\Lambda_{\rm b}$ baryon \cite{Tsushima:2002ua,Tsushima:2002sm}, $\Sigma_{\rm b}$ baryon, $\bar{B}$ meson, $B$ meson \cite{Hilger:2008jg,Yasui:2012rw} for open bottom and $\Upsilon$ for hidden bottom.
Some of them were motivated by the possible attractive force between a charm (bottom) hadron and a nucleon which was found in studies of hadronic molecules as exotic hadrons \cite{Yasui:2009bz,Yamaguchi:2011xb,Yamaguchi:2011qw,Liu:2011xc,Meguro:2011nr,Ohkoda:2011vj,Ohkoda:2012hv} (see also \cite{Swanson:2006st,Voloshin:2007dx,Nielsen:2009uh,Brambilla:2010cs}).
The results of those studies suggest that a charm (bottom) hadron 
 can be bound in nuclear matter.
Such systems are interesting, because they will help us to study, not only (i) the interaction between the heavy hadron and nucleon, but also (ii) the change of the properties of the heavy hadron in nuclear medium and (iii) the change of nuclear medium (including the partial restoration of the dynamical breaking of chiral symmetry) caused by the heavy hadron as an impurity.
The exotic nuclear systems with heavy flavors can be studied experimentally in J-PARC, GSI-FAIR and so on.
The study of heavy quarks in quark matter is also interesting, when the state at high density and low temperature can be produced in heavy ion collisions.
Actually it is expected that there are rich structures like color superconductivity in quark matter \cite{Alford:2007xm,Fukushima:2010bq}.

In the systems with a single heavy meson or quark, the heavy mass limit is very useful for analysis of their properties in vacuum \cite{Manohar:2000dt,Grozin:2004yc,Burdman:1992gh,Wise:1992hn,Yan:1992gz,Casalbuoni:1996pg}.
In the present article, let us study the dynamics of the heavy meson or quark with a single heavy flavor, which is embedded in finite density medium at zero temperature, and investigate the behavior in the large limit of their masses in the medium.
As systems, we consider a $\bar{D}$ ($B$) meson embedded in nuclear matter and a charm (bottom) quark embedded in quark matter.
Although the former and latter systems are quite different, both of them exhibit a similar behavior in the large mass limit as discussed below.
We note that, in our systems, the fermions composing the Fermi surface (nucleons in nuclear matter or light quarks in quark matter) and the heavy impurity particle ($\bar{D}$ ($B$) meson or charm (bottom) quark) belong to the fundamental representation of SU($n$) symmetry with an integer $n \ge 2$. 
In fact, a $\bar{D}$ ($B$) meson as well as nucleons are doublet states in SU(2) isospin symmetry, and  a charm (bottom) quark as well as light quarks are triplet states in SU(3) color symmetry.
In the present discussion, we suppose that SU($n$) symmetry is a global symmetry, not only for isospin symmetry, but also for color symmetry.
For generality of the formulation, we consider that the fermions and the heavy impurity particle belong to the fundamental representation of SU($n$) symmetry (isospin  for $n=2$ and color for $n=3$).
We furthermore suppose a vector current interaction between the fermion and the heavy impurity particle with a small coupling constant so that we can apply the perturbative calculation.
Such a simple interaction can be used as far as the low energy region near the Fermi surface is concerned.
With this setup, we discuss the scattering of the fermions (nucleons or light quarks) and the heavy impurity particle ($\bar{D}$ ($B$) meson or charm (bottom) quark).

In condensed matter physics, such a situation has been known as ``Kondo problem."
There, spin-half electrons in conduction band are coupled to an impurity atom with a non-zero (pseudo)spin through the ``(pseudo)spin-spin" interaction.
In 1964, Kondo found that the interaction between the conduction electrons and the impurity atom causes a logarithmic enhancement by the temperature in the system or the energy of the scattering fermions when the one-loop scattering amplitude is considered \cite{Kondo:1964}.
The problem was analyzed in detail in the scaling method, and was further developed by the numerical renormalization group analysis and so on \cite{Hewson}.
Throughout those studies, it was recognized that, even though the coupling constant is small, the quantum fluctuation from creations of particles and holes near the Fermi surface enhances the scattering amplitude with higher order loops, and the perturbation breaks down in the limit of low temperature or small scattering energy.

An important comment is in order.
In the original work by Kondo, it was assumed tacitly
 that the impurity atom has an infinitely heavy mass.
This assumption will be acceptable for electron-atom systems, because of the small mass ratio 
 between electron and atom.
However, it is not necessarily the case that such an assumption can also be applied to $\bar{D}$ ($B$) meson in nuclear medium and charm (bottom) quark in quark matter.
It is rather a nontrivial problem how 
the large mass of the impurity particle plays a role in the medium.
The present study is mainly devoted to this problem. 
Actually, this is an important problem in the hadron physics in order to understand how the properties of heavy mesons, such as $\bar{D}$ and $D$ ($B$ and $\bar{B}$) mesons, in nuclear matter are different from those of light mesons, such as $K$ and $\bar{K}$ mesons, in nuclear matter \cite{Akaishi:2002bg,Hyodo:2011ur}.
In the former the heavy mass limit can be applied, while in the latter it cannot.
It will be also important to study the difference of charm (bottom) quark and strange quark in quark matter.
As a matter of fact, this paper covers the result given by Kondo. 

The paper is organized as followings.
In section \ref{sec:boson}, we discuss the heavy impurity boson embedded in the Fermi gas (e.g. a $\bar{D}$ ($B$) meson in nuclear matter),
and derive the scattering amplitude for the heavy impurity boson and the fermion composing the Fermi surface. 
In section \ref{sec:fermion}, we discuss the heavy impurity fermion embedded in the Fermi gas (e.g. a charm (bottom) quark in quark matter).
In both two cases, we suppose a vector current interaction with a small coupling constant between the fermion and the heavy impurity particle, 
and analyze the scattering amplitude for each heavy impurity particle up to one-loop level.
Then we show that the scattering amplitudes have a logarithmic enhancement in the heavy mass limit, and the perturbation breaks down for any small coupling constant.
In section \ref{sec:discussion}, we discuss the related topics, and in the final section we summarize the discussion and give perspectives.

\section{Heavy impurity boson} \label{sec:boson}

We consider a $\bar{D}$ ($B$) meson embedded in nuclear matter.
As a model for the interaction with SU(2) isospin symmetry between a $\bar{D}$ ($B$) meson and a nucleon, we consider the vector current interaction with an isospin factor of $\vec{\lambda}_{\rm f} \cdot \vec{\lambda}_{\rm B}$ (see below) and a small coupling constant, and analyze the scattering amplitude up to one-loop level.
To see how the internal degrees of freedom works in the scattering amplitude, we generally extend SU(2) symmetry to SU($n$) symmetry with an arbitrary integer $n \ge 2$.

\subsection{Interaction with SU($n$) symmetry}

We consider the interaction Lagrangian given by the vector current interaction with SU($n$) symmetry for the fermion and the heavy impurity boson
\begin{eqnarray}
{\cal L}_{\rm{B,int}} = -\frac{G_{\rm B}}{2} \sum_{j=1}^{n^2-1} \left( \bar{\psi} \gamma_{\mu} \lambda^{j}_{\rm f} \psi \right) \left( -i \partial^{\mu} \Phi^{\dag} \lambda^{j}_{\rm B} \Phi +  \Phi^{\dag} \lambda^{j}_{\rm B} i\partial^{\mu} \Phi \right),
\label{eq:int_boson_0}
\end{eqnarray}
where the fermion field $\psi = (\psi_{1}, \cdots, \psi_{n})^{\rm t}$ and the heavy impurity boson field $\Phi = (\Phi_{1}, \cdots, \Phi_{n})^{\rm t}$ belong to the fundamental representation of SU($n$) symmetry.
The $n \times n$ matrices $\lambda^{j}_{\rm f}/2$ and $\lambda^{j}_{\rm B}/2$ ($j=1$, $\cdots$, $n^2-1$) are the generators of SU($n$) symmetry for the fermion and the heavy impurity boson, respectively.
The coupling constant $G_{\rm B}$ is assumed to be a small number for which the perturbation can be applied. 
We suppose that the initial state of the heavy impurity boson has a four-dimensional momentum $P^{\mu}$ and the final state of the heavy impurity boson has $P^{\prime\,\mu}$, and rewrite the interaction Lagrangian in momentum space.
Considering that the mass $M_{\rm B}$  of the heavy impurity boson is very large, we decompose $P^{\mu}$ and $P^{\prime\,\mu}$ as
\begin{eqnarray}
P^{\mu} &=& M_{\rm B} v^{\mu} + k^{\mu},
\label{eq:momentum_decomposition} \\
P^{\prime\,\mu} &=& M_{\rm B} v^{\mu} + k^{\prime\,\mu},
\end{eqnarray}
where 
$v^{\mu}$ is a four-dimensional velocity with an on-mass-shell condition $v^2 = 1$, and $k$ and $k^{\prime}$ are off-mass-shell (residual) momenta.
When we ignore the terms with the residual momentum, which are suppressed by the order of $1/M_{\rm B}$ from ones with $v^{\mu}$, the above Lagrangian is rewritten as
\begin{eqnarray}
{\cal L}_{\rm{B,int}} = -G_{\rm B}M_{\rm B}  \sum_{j=1}^{n^2-1} \left( \bar{\psi} \gamma_{\mu} \lambda^{j}_{\rm f} \psi \right) \left( \Phi^{\dag} v^{\mu} \lambda^{j}_{\rm B} \Phi + {\cal O}(k/M_{\rm B},k'/M_{\rm B}) \right).
\label{eq:int_boson}
\end{eqnarray}
To drop the terms at ${\cal O}(k/M_{\rm B},k'/M_{\rm B})$ will be justified, because the large mass limit for the heavy impurity boson is adopted in the discussion.
We assume that the heavy impurity boson is at rest in the medium and set $v^{\mu}=(1, \vec{0}\,)$.

\subsection{Scattering amplitude}

\begin{figure}[tbp]
\includegraphics[width=4cm,angle=0]{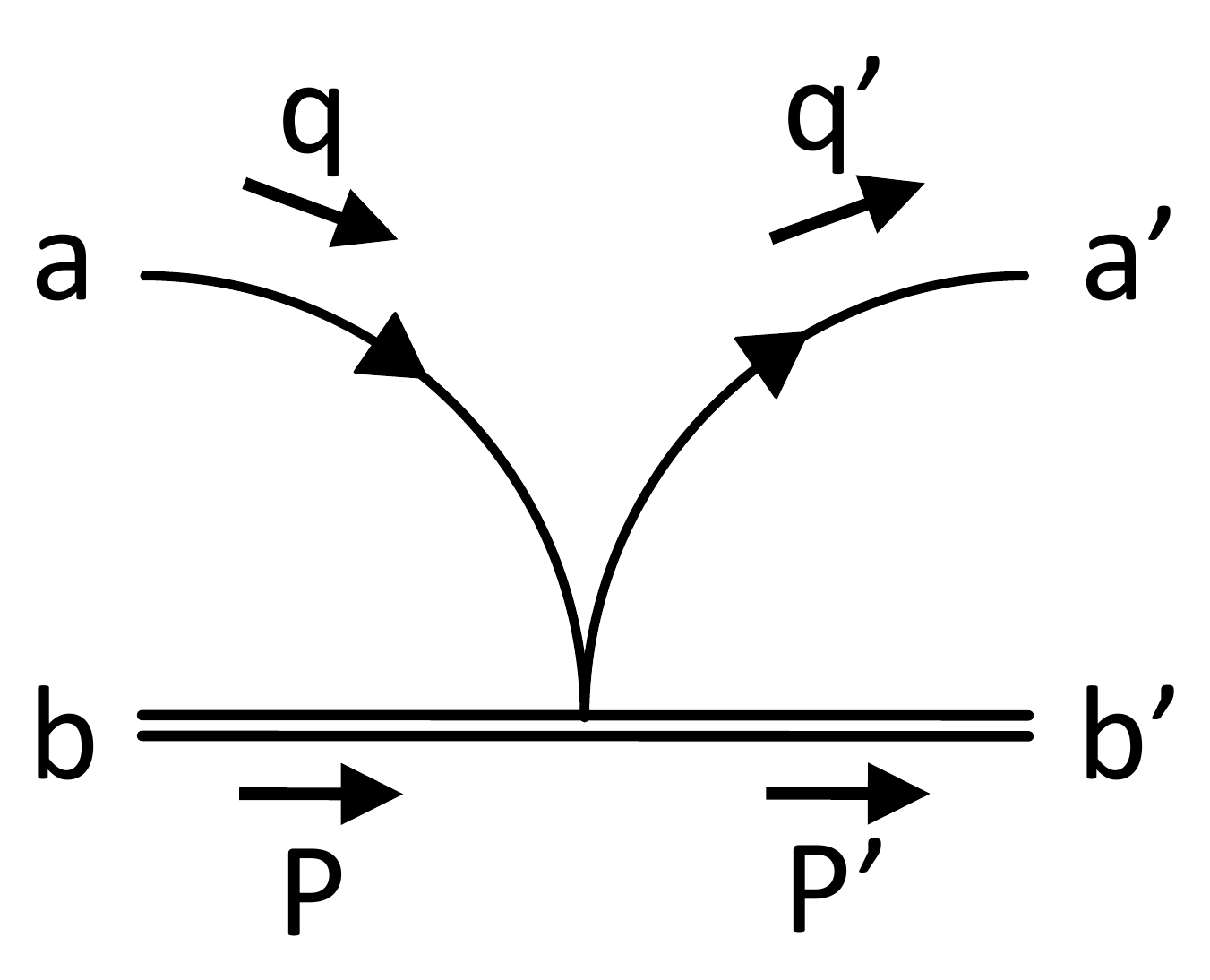}
\caption{The diagram of the first order scattering amplitude (the Borne term) of the fermion $\psi_{a(a')}$ ($a,a'=1,\cdots,n$) (single line) composing the Fermi surface and the heavy impurity boson $\Phi_{b(b')}$ ($b,b'=1,\cdots,n$) (double line) embedded in the Fermi gas is shown. $q$ ($q'$) is the initial (final) four-dimensional momentum of the fermion, and $P$ ($P'$) is the initial (final) four-dimensional momentum of the boson.}
\label{fig:Fig1}
\end{figure}
\begin{figure}[tbp]
\includegraphics[width=7cm,angle=0]{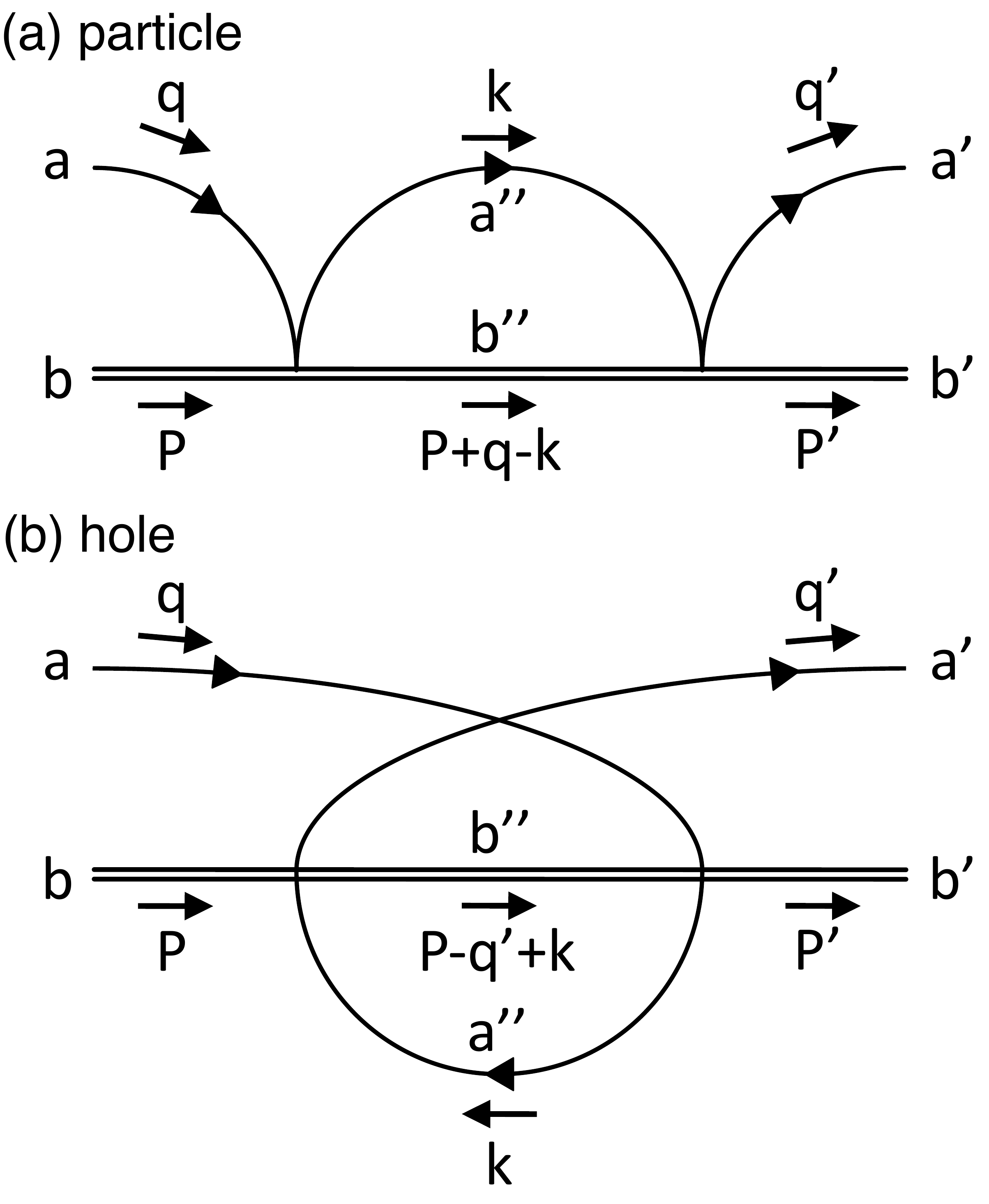}
\caption{The diagrams of the second order scattering amplitude of the fermion and the heavy impurity boson are shown.  $k$ is the internal moment in the loop. The other notations are common to that in Fig.~\ref{fig:Fig1}.}
\label{fig:Fig2}
\end{figure}

Based on the above interaction Lagrangian, we consider the scattering amplitude for the fermion and the heavy impurity boson up to one-loop level as shown in Figs.~\ref{fig:Fig1} and \ref{fig:Fig2}.
We introduce the initial (final) three-dimensional momentum $\vec{q}$ ($\vec{q}\,'$) of the fermion which is very close to the Fermi surface, and the initial (final) momentum  $\vec{P}$ ($\vec{P}\,'$) of the heavy impurity boson with $\vec{P}=0$ at rest in the medium.
Then, it will be induced that the initial and final momenta of the scattering fermion are same ($\vec{q}\,'\simeq \vec{q}$\,) and the recoil of the heavy impurity boson is negligible ($\vec{P}\,'\simeq 0$).
Indeed, from the momentum conservation $\vec{q}+\vec{P} = \vec{q}\,'+\vec{P}\,'$ and the energy conservation $\vec{q}^{\,\,2}/2m+\vec{P}^{\,2}/2M_{\rm B} = \vec{q}\,'^{\,2}/2m+\vec{P}\,'^{\,2}/2M_{\rm B}$ for the nonrelativistic fermion, we obtain the desired relation under the conditions $|\vec{q}\,| \simeq k_{\rm F}$, $\vec{P}=0$ and $|\vec{q}\,'| \ge k_{\rm F}$, where the last condition is given by the Pauli blocking effect
 inside the Fermi sphere with Fermi momentum $k_{\rm F}$.
 It is also the case for the relativistic fermion.
Hence, we use $\vec{q}\,'=\vec{q}$ and $\vec{P}=\vec{P}\,'=0$ throughout the present discussion.
The components in the fundamental representation of SU($n$) for the fermion and the heavy impurity boson are denoted by $a$ and $b$ ($a'$ and $b'$) for the initial (final) states, respectively.

To begin with, the first order contribution (the Born term) in the scattering amplitude in Fig.~\ref{fig:Fig1} is obtained as
\begin{eqnarray}
-i {\cal M}^{(1)}_{\rm B} = -i G_{\rm B}M_{\rm B} \, \bar{u}_{q,a'} v\hspace{-0.5em}/ u_{q,a} \, (\vec{\lambda}_{\rm f})_{a'a} \! \cdot \! (\vec{\lambda}_{\rm B})_{b'b}, 
\label{eq:amp1_boson}
\end{eqnarray}
with the four-dimensional velocity $v^{\mu}=(1,\vec{0}\,)$ for the heavy impurity boson and the four component spinor $u_{q}$ ($\bar{u}_{q}$) for the initial (final) fermion with momentum $q=(q_{0}, \vec{q}\,)$.
We note again that the absolute value of the three-dimensional momentum $\vec{q}$ should be close to the Fermi momentum; $|\vec{q}\,| \simeq k_{\rm F}$.
Concerning the coupling in the interaction, we neglect the higher order terms in expansion by power of $1/M_{\rm B}$.
Apparently, the first order scattering amplitude has no singular behavior in the large mass limit of $M_{\rm B}$.


Next, we consider the second order contribution which contains the fermion propagator in the loop.
In the fermion propagator in finite density medium at zero temperature, we use the in-medium fermion propagator,
\begin{eqnarray}
 (p\hspace{-0.4em}/+m) \left[ \frac{i}{p^{2}-m^{2}+i\varepsilon} - 2\pi \delta(p^{2}-m^{2}) \theta(p_{0}) \theta(k_{\mathrm{F}}-|\vec{p}\,|) \right] \mathbf{1}_{n \times n},
 \label{eq:propagator}
\end{eqnarray}
with the fermion mass $m$, the four-dimensional momentum $p^{\mu}=(p_{0}, \vec{p}\,)$, and the Fermi momentum $k_{\mathrm{F}}$ \cite{Kaiser:2001jx}.
Here $\varepsilon$ is an infinitely small positive number and $\mathbf{1}_{n \times n}$ is an $n \times n$ unit matrix corresponding to SU($n$) symmetry.
The second term with the delta function and the step functions indicates the Pauli blocking effect for the fermions in the Fermi sphere.
Because the fermions with positive energy are occupied up to the Fermi surface in the momentum space, the on-mass-shell fermions in the Fermi sphere cannot propagate. 

By using the in-medium fermion propagator Eq.~(\ref{eq:propagator}), let us consider the second order scattering amplitude in Fig.~\ref{fig:Fig2}.
Before going to the precise calculation, it will be worthwhile to see na\"ively how the scattering amplitude would behave in large $M_{\rm B}$.
Because the strength of the interaction vertex is proportional to $G_{\rm B} M_{\rm B}$ in Eq.~(\ref{eq:int_boson}), the first order scattering amplitude is also proportional to $G_{\rm B} M_{\rm B}$ as obtained above.
Concerning the second order contribution, we pay an attention to that the heavy impurity boson propagator is proportional to $1/M_{\rm B}$ in large $M_{\rm B}$, because $(P^{2}-M_{\rm B}^{2})^{-1}=((M_{\rm B}v+k)^{2}-M_{\rm B}^{2})^{-1} \simeq (2M_{\rm B} v \cdot k)^{-1}$ from Eq.~(\ref{eq:momentum_decomposition}), as far as the heavy impurity boson is close to the on-mass-shell state.
Hence we might observe that the second order scattering amplitude will be proportional to $(G_{\rm B}M_{\rm B})^2 ({\rm{vertex}}) \times M_{\rm B}^{-1} ({\rm{propagator}}) = G_{\rm B}^2 M_{\rm B}$, when the momentum cutoff in the loop is fixed.
Then, because the power of $M_{\rm B}$ is the same in the first and second order scattering amplitudes, it seems for any large value of $M_{\rm B}$ that the perturbation is applicable for the small number of $G_{\rm B}$.
However, we will present that this na\"ive analysis does not hold for the interaction Eq.~(\ref{eq:int_boson_0}) in finite density medium at zero temperature.

The precise form of the second order scattering amplitude is given as
\begin{eqnarray}
-i{\cal M}^{(2)}_{\rm B}=-i{\cal M}^{(2)}_{\rm B} ({\rm{Fig.~\ref{fig:Fig2} (a)}})-i{\cal M}^{(2)}_{\rm B} ({\rm{Fig.~\ref{fig:Fig2} (b)}}),
\end{eqnarray}
where each contribution is
\begin{eqnarray}
 -i {\cal M}^{(2)}_{\rm B} ({\rm{Fig.~\ref{fig:Fig2} (a)}}) &=& \left(-iG_{\rm B}M_{\rm B}\right)^{2} \left\{ 4\left(1-\frac{1}{n^2}\right) \delta_{a'a} \delta_{b'b} +\left( -\frac{4}{n} \right) (\vec{\lambda}_{\rm f})_{a'a} \! \cdot \! (\vec{\lambda}_{\rm B})_{b'b} \right\}  \nonumber \\
&& \times \bar{u}_{q',a'} v\hspace{-0.5em}/ \int \frac{{\rm d}^{4}k}{(2\pi)^4} ( k\hspace{-0.5em}/ + m )  
 \left[ \frac{i}{k^2-m^2+i\varepsilon} - 2\pi \delta(k^2-m^2) \theta(k_{0}) \theta(k_{\rm F}-|\vec{k}\,|) \right] \nonumber \\
&& \times \frac{i}{(P+q-k)^2-M_{\rm B}^2+i\varepsilon} \, v\hspace{-0.5em}/ u_{q,a},
\label{eq:boson_fig2a}
\end{eqnarray}
in Fig.~\ref{fig:Fig2} (a), and 
\begin{eqnarray}
 -i {\cal M}^{(2)}_{\rm B} ({\rm{Fig.~\ref{fig:Fig2} (b)}}) &=& \left(-iG_{\rm B}M_{\rm B}\right)^{2} \left\{ 4\left(1-\frac{1}{n^2}\right) \delta_{a'a} \delta_{b'b} + 2 \left( n-\frac{2}{n} \right) (\vec{\lambda}_{\rm f})_{a'a} \! \cdot \! (\vec{\lambda}_{\rm B})_{b'b} \right\}  \nonumber \\
&& \times \bar{u}_{q,a'} v\hspace{-0.5em}/ \int \frac{{\rm d}^{4}k}{(2\pi)^4} ( k\hspace{-0.5em}/ + m )  
 \left[ \frac{i}{k^2-m^2+i\varepsilon} - 2\pi \delta(k^2-m^2) \theta(k_{0}) \theta(k_{\rm F}-|\vec{k}\,|) \right] \nonumber \\
&& \times \frac{i}{(P-q+k)^2-M_{\rm B}^2+i\varepsilon} \, v\hspace{-0.5em}/ u_{q,a},
\label{eq:boson_fig2b}
\end{eqnarray}
in Fig.~\ref{fig:Fig2} (b).
$k$ is the internal momentum in the loop.
To obtain the above equations, we use the identities
\begin{eqnarray}
&& \sum_{a'', b''=1}^{n} \sum_{i,j=1}^{n^2-1} \left( \lambda^{i}_{\rm f} \right)_{a'a''} \left( \lambda^{i}_{\rm B} \right)_{b'b''} \left( \lambda^{j}_{\rm f} \right)_{a''a} \left( \lambda^{j}_{\rm B} \right)_{b''b} \nonumber \\
&=& 4 \left( 1-\frac{1}{n^2} \right) \delta_{a'a} \delta_{b'b} + \left( -\frac{4}{n} \right) \sum_{i=1}^{n^2-1} \left( \lambda^{i}_{\rm f} \right)_{a'a} \left( \lambda^{i}_{\rm B} \right)_{b'b},
\label{eq:identity1}
\end{eqnarray}
and
\begin{eqnarray}
&& \sum_{a'', b''=1}^{n} \sum_{i,j=1}^{n^2-1} \left( \lambda^{i}_{\rm f} \right)_{a'a''} \left( \lambda^{j}_{\rm B} \right)_{b'b''} \left( \lambda^{j}_{\rm f} \right)_{a''a} \left( \lambda^{i}_{\rm B} \right)_{b''b} \nonumber \\
&=& 4 \left( 1-\frac{1}{n^2} \right) \delta_{a'a} \delta_{b'b} + 2 \left( n-\frac{2}{n} \right) \sum_{i=1}^{n^2-1} \left( \lambda^{i}_{\rm f} \right)_{a'a} \left( \lambda^{i}_{\rm B} \right)_{b'b},
\label{eq:identity2}
\end{eqnarray}
in Eqs.~(\ref{eq:boson_fig2a}) and $(\ref{eq:boson_fig2b})$, respectively.
It is important in the following discussion that the sign of the coefficient $-4/n$ in the factor $(\vec{\lambda}_{\rm f})_{a'a} \cdot (\vec{\lambda}_{\rm B})_{b'b}$ in Eq.~(\ref{eq:identity1}) are opposite to that of $2 ( n-2/n )$ in the corresponding factor in Eq.~(\ref{eq:identity2}).
After integrating by $k_{0}$ in the integrals, then, we obtain the result for the finite density part
\begin{eqnarray}
-i {\cal M}^{(2)}_{\rm B}
&=&
\left(-iG_{\rm B}M_{\rm B}\right)^{2}
4\left(1-\frac{1}{n^2}\right) \delta_{a'a} \delta_{b'b}
 \nonumber \\
&& \times \bar{u}_{q,a'} v\hspace{-0.5em}/ \left[
i \int_{|\vec{k}\,| \ge k_{\rm F}} \frac{{\rm d}^{3}\vec{k}}{(2\pi)^3} \frac{1}{2\epsilon_{k}}
 \frac{ \epsilon_{k}\gamma_{0} - \vec{k} \! \cdot \! \vec{\gamma} + m }{(P_0+q_0-\epsilon_k)^2-E_{P+q-k}^2+i\varepsilon} \right. \nonumber \\
 && \left. 
 +(-i) \int_{|\vec{k}\,| \le k_{\rm F}} \frac{{\rm d}^{3}\vec{k}}{(2\pi)^3} \frac{1}{2\epsilon_{k}}
 \frac{ \epsilon_{k}\gamma_{0} - \vec{k} \! \cdot \! \vec{\gamma} + m }{(P_0-q_0+\epsilon_k)^2-E_{P-q+k}^2+i\varepsilon}
\right] \, v\hspace{-0.5em}/ u_{q,a} \nonumber \\
&&+ \left(-iG_{\rm B}M_{\rm B}\right)^{2}
(\vec{\lambda}_{\rm f})_{a'a} \! \cdot \! (\vec{\lambda}_{\rm B})_{b'b}
  \nonumber \\
&& \times \bar{u}_{q,a'}v\hspace{-0.5em}/ \left[
\left( -\frac{4}{n} \right) i \int_{|\vec{k}\,| \ge k_{\rm F}} \frac{{\rm d}^{3}\vec{k}}{(2\pi)^3} \frac{1}{2\epsilon_{k}}
 \frac{ \epsilon_{k}\gamma_{0} - \vec{k} \! \cdot \! \vec{\gamma} + m }{(P_0+q_0-\epsilon_k)^2-E_{P+q-k}^2+i\varepsilon} 
 \right. \nonumber \\
&& \left. + 2\left( 2-\frac{2}{n} \right) (-i) \int_{|\vec{k}\,| \le k_{\rm F}} \frac{{\rm d}^{3}\vec{k}}{(2\pi)^3} \frac{1}{2\epsilon_{k}} 
 \frac{ \epsilon_{k}\gamma_{0} - \vec{k} \! \cdot \! \vec{\gamma} + m }{(P_0-q_0+\epsilon_k)^2-E_{P-q+k}^2+i\varepsilon}
\right] \, v\hspace{-0.5em}/ u_{q,a},
\label{eq:amp2_boson}
\end{eqnarray}
where we define $P_{0}=E_{P}$ and $q_{0}=\epsilon_{q}$ with $E_{P}=\sqrt{\vec{P}\,^2+M_{\rm B}^{2}}$ for the energy of the heavy impurity boson and $\epsilon_{q}=\sqrt{\vec{q}\,^2+m^{2}}$ for the energy of the scattering fermion.
We leave only the contribution from finite density, because we are interested in the low energy region around the Fermi surface.
In Eq.~(\ref{eq:amp2_boson}), the former integral with $|\vec{k}|>k_{\rm F}$ (the latter one with $|\vec{k}|<k_{\rm F}$) in each squared bracket is the contribution from the particles  (holes) in the loop shown in Fig.~\ref{fig:Fig2} (a) ((b)).
For convergence of the momentum integrals, we introduce the momentum cutoffs $\Lambda_{\rm high}$ for $|\vec{k}|>k_{\rm F}$ and $\Lambda_{\rm low}$ for $|\vec{k}|<k_{\rm F}$.
Finally, the scattering amplitude up to one-loop level is given by
\begin{eqnarray}
-i{\cal M}_{\rm B} = -i{\cal M}^{(1)}_{\rm B} -i{\cal M}^{(2)}_{\rm B}.
\label{eq:amp_tot}
\end{eqnarray}
From now on, our effort is devoted to analyze the behavior of the second order amplitude in the large mass limit of the heavy impurity boson.

\subsection{Large mass limit of heavy impurity boson}

We consider the large mass limit for the heavy impurity boson ($M_{\rm B} \rightarrow \infty$).
We expand $E_{P}$ by $1/M_{\rm B}$ as $E_{P} \simeq M_{\rm B} + \vec{P}\,^{2}/2M_{\rm B}$ and
consider only the leading term by neglecting the higher order contributions.
In the following, for the sake of the simple presentation, we introduce new dimensionless variables $\eta$ and $\Lambda$, which are defined by
$|\vec{q}\,| = k_{\rm F} (1+\eta)$
for the three-dimensional momentum $\vec{q}$ of the scattering fermion, and
$\Lambda_{\rm high} = k_{\rm F} (1+\Lambda)$ and
$\Lambda_{\rm low} = k_{\rm F} (1-\Lambda)$
for the cutoff parameters $\Lambda_{\rm high}$ and $\Lambda_{\rm low}$.
We note that $\eta \ge 0$ is a nonnegative small number because the initial state of the scattering fermion is supposed to lie on or outside of the Fermi sphere.
We also note that the cutoff parameter $\Lambda$ is fixed as a finite number ( $0 < \Lambda < 1$).
Because the result of the integrals in Eq.~(\ref{eq:amp2_boson}) is lengthy, 
we simplify the analysis 
 by considering the two limiting cases for the scattering fermions; nonrelativistic fermions and relativistic fermions.
Although the nonrelativistic fermions are much realistic in nuclear matter, the relativistic fermions will be found to be useful for our analysis.

\subsubsection{Nonrelativistic fermions}
We consider the nonrelativistic fermions.
We assume that the mass of the scattering fermion is larger than the momentum ($\simeq k_{\rm F}$), and expand $\epsilon_{k}$ as $\epsilon_{k} \simeq m + \vec{k}\,^2/2m$.
Concerning the heavy impurity boson, we keep the mass of the heavy impurity boson still much heavier than the mass of the scattering fermion, and consider  $m/M_{\rm B}$ as a small number; $m/M_{\rm B} \ll 1$.

To begin the discussion, we assume that the initial state of the scattering fermion is in the ground state and its momentum lies on the Fermi surface; $|\vec{q}\,|=k_{F}$ ($\eta=0$).
Then, we take the limit $M_{\rm B} \rightarrow \infty$ ($m/M_{\rm B} \rightarrow 0$ with keeping $m$ to be a constant), and obtain the leading contribution (see also Appendix~\ref{sec:calc})
\begin{eqnarray}
&& \lim_{M_{\rm B} \rightarrow \infty} \lim_{\eta \rightarrow 0} \left( -i {\cal M}^{(2)}_{\rm B} \right)_{\rm{nonrel}}
\label{eq:amp2_boson_non-rela} \\
&=&  \left( -iG_{\rm B}M_{\rm B} \right)^2 \delta_{a'a} \delta_{b'b} \nonumber \\ 
&& \times i  \frac{1}{2M_{\rm B}} \frac{2mk_{\rm F}}{4\pi^{2}} 4 \left( 1-\frac{1}{n^2} \right) \left[ -4\Lambda + \log \left( 1+\frac{\Lambda}{2} \right) - \log \left( 1-\frac{\Lambda}{2} \right) \right] \bar{u}_{q,a'} \frac{1+v\hspace{-0.5em}/}{2} u_{q,a} \nonumber \\
&& + \left( -iG_{\rm B}M_{\rm B} \right)^2 (\vec{\lambda}_{\rm f})_{a'a} \cdot (\vec{\lambda}_{\rm B})_{b'b} \nonumber \\
&& \times i \frac{1}{2M_{\rm B}} \frac{2mk_{\rm F}}{4\pi^{2}} \left[ -4\left( n-\frac{4}{n} \right) \Lambda + 2n \log \Lambda + \left(-\frac{4}{n}\right) \log \left( 1+\frac{\Lambda}{2} \right) - 2 \left( n-\frac{2}{n} \right) \log \left( 1-\frac{\Lambda}{2} \right) \right. \nonumber \\
&& \left.- 2n \log \frac{m}{M_{\rm B}} +2n (1-\log2 )\right] \bar{u}_{q,a'} \frac{1+v\hspace{-0.5em}/}{2} u_{q,a}. \nonumber
\end{eqnarray}
The spinor for nonrelativistic fermion is defined as $u_{q} \simeq \sqrt{2m} (\chi,0)^{\rm t}$
with two-component spinor $\chi$.

In the above result, we find the logarithmic behavior of $\log M_{\rm B}$.
To see the importance of this term, let us compare the first order scattering amplitude Eq.~(\ref{eq:amp1_boson}) with the second order scattering amplitude Eq.~(\ref{eq:amp2_boson_non-rela}).
We notice that the first order scattering amplitude contains the factor $G_{\rm B}M_{\rm B}$, while the second order scattering  amplitude contains 
not only $G_{\rm B}^{\,2} M_{\rm B}$ but also $G_{\rm B}^{\,2} M_{\rm B} \log M_{\rm B}$.
When we consider the heavy mass limit ($M_{\rm B} \rightarrow \infty$), due to the presence of $\log M_{\rm B}$, the second order scattering amplitude overcomes the first order one.
Therefore, 
the perturbative calculation is not applicable
for any small coupling between the fermion and heavy impurity boson in the heavy mass limit,
and the system becomes a strongly interacting one.
The critical mass of $M_{\rm B}$ which gives the second order scattering amplitude comparable with the first order one is obtained as
\begin{eqnarray}
 M_{\rm B}^{\rm{cr,nonrel}} = m \exp \left( \frac{1}{n |G_{\rm B}|} \frac{4\pi^{2}}{2mk_{\rm F}} \right),
 \label{eq:critical_mass_boson_nonrela}
\end{eqnarray}
when Eq.~(\ref{eq:amp1_boson}) and the term proportional to $\log m/M_{\rm B}$ in Eq.~(\ref{eq:amp2_boson_non-rela}) are compared.
For $M_{\rm B} \gg M_{\rm B}^{\rm{cr,nonrel}}$, the second order scattering amplitude is larger than the first order one, and the perturbation breaks down.
We here note that the value of $M_{\rm B}^{\rm{cr,nonrel}}$ should not be considered seriously because it just gives an order of the critical mass.

In our discussion, 
the factor $\vec{\lambda}_{\rm f} \cdot \vec{\lambda}_{\rm B}$
  in the interaction Eq.~(\ref{eq:int_boson_0}) plays a significantly important role.
Indeed the opposite sign in the coefficients, $-4/n$ and $2(2-2/n)$, in the terms with $(\vec{\lambda}_{\rm f}\,)_{a'a} \!\cdot\! (\vec{\lambda}_{\rm B}\,)_{b'b}$ in Eq.~(\ref{eq:amp2_boson}) induces the factor $\log M_{\rm B}$.
On the other hand, in the terms with $\delta_{a'a}\delta_{b'b}$ in Eq.~(\ref{eq:amp2_boson}), we confirm that 
 no logarithmic term exists. 
We comment that the Fermi surface is also important, because we find no logarithmic term in the vacuum part in Eqs.~(\ref{eq:boson_fig2a}) and (\ref{eq:boson_fig2b}).

We should emphasize that the logarithmic behavior of $\log M_{\rm B}$ in the scattering amplitude appears only in the limit of large $M_{\rm B}$.
Otherwise, there is no such singularity.
To see this, let us consider the case that the mass of the heavy impurity boson is not so large.
Supposing that the heavy impurity boson mass happens to be equal to the fermion mass, $M_{\rm B}=m$, we analyze the loop integrals in Eq.~(\ref{eq:amp2_boson}) and obtain
\begin{eqnarray}
&& \lim_{M_{\rm B} \rightarrow m} \lim_{\eta \rightarrow 0} \left( -i{\cal M}^{(2)}_{\rm B} \right)_{\rm{nonrel}} \\
\label{eq:amp2_boson_non-rela_2}
&=&  \left( -iG_{\rm B}M_{\rm B} \right)^2  \delta_{a'a} \delta_{b'b} \nonumber \\ 
&& \times i \frac{1}{2M_{\rm B}} \frac{k_{\rm F} \Lambda}{4\pi^{2}} 4 \left( 1-\frac{1}{n^2} \right) \left[ -\left(1+\frac{\Lambda}{2}\right) \frac{1}{2} \log \frac{2+\Lambda}{\Lambda} + \left( 1-\frac{\Lambda}{2} \right) \frac{1}{2} \log \frac{2-\Lambda}{\Lambda} \right] \nonumber \\
&& \times \bar{u}_{q,a'} \frac{1+v\hspace{-0.5em}/}{2} u_{q,a} \nonumber \\
&& + \left( -iG_{\rm B}M_{\rm B} \right)^2 (\vec{\lambda}_{\rm f})_{a'a} \cdot (\vec{\lambda}_{\rm B})_{b'b} \nonumber \\
&& \times i \frac{1}{2M_{\rm B}} \frac{k_{\rm F}\Lambda}{4\pi^{2}} \left[ n - \left(-\frac{4}{n} \right) \left( 1+ \frac{\Lambda}{2} \right) \frac{1}{2} \log \frac{2+\Lambda}{\Lambda} + 2\left( n-\frac{2}{n} \right) 2 \left( 1-\frac{\Lambda}{2} \right) \frac{1}{2} \log\frac{2-\Lambda}{\Lambda} \right] \nonumber\\ 
&& \times \bar{u}_{q,a'} \frac{1+v\hspace{-0.5em}/}{2} u_{q,a}, \nonumber
\end{eqnarray}
for the fermion on the Fermi surface.
We confirm that there is no logarithmic term for $M_{\rm B}=m$.
Thus, for presence of $\log M_{\rm B}$, it is important that $M_{\rm B}$ is much larger than other scales in the system.

As a brief summary in the analysis for the nonrelativistic fermions, starting from the interaction in Eq.~(\ref{eq:int_boson_0}), we have shown that
the perturbation breaks down for any small coupling constant in the limit of large $M_{\rm B}$.
It should be noticed that, when the interaction does not contain derivatives for the heavy impurity boson field, like a scalar interaction,
the above conclusion does not hold.
This is because in the scalar interaction the first order scattering amplitude is proportional $G_{\rm B}$, and the second order one contains $G_{\rm B}^{\,2}/M_{\rm B}$ or $(G_{\rm B}^{\,2}/M_{\rm B}) \log M_{\rm B}$ (the same convention of the coupling constant $G_{\rm B}$ is used in the scalar interaction).
Hence, the second order contribution can become smaller than the first order one in the limit of large $M_{\rm B}$, and the perturbation can be applied for the small coupling constant $G_{\rm B}$.

\subsubsection{Relativistic fermions}

Let us move to the discussion for the relativistic fermions.
For simplicity, we consider the massless fermions.
We will see that the similar conclusion holds in the large mass limit of the heavy impurity boson.
We note that, in this case, the expansion parameter cannot be given by the ratio of the fermion mass and the heavy impurity boson mass, $m/M_{\rm B}$, because the fermions now are massless.
Instead, we use the ratio of the Fermi momentum and the heavy impurity boson mass, $k_{\rm F}/M_{\rm B}$, for expansion parameter for large $M_{\rm B}$ with finite $k_{\rm F}$.

The first order contribution of the scattering amplitude is given by Eq.~(\ref{eq:amp1_boson}), provided that the nonrelativistic spinor is replaced to the relativistic spinor
$u_{q} = \sqrt{E_{q}} (\chi, \vec{\sigma} \! \cdot \! \hat{q} \, \chi  )^{\rm t}$
with the two component spinor $\chi$, the energy $E_{q}=|\vec{q}\,|$ and the unit vector $\hat{q} = \vec{q}/|\vec{q}\,|$ in $\vec{q}$ direction.

For the second order contribution, 
for the scattering fermion on the Fermi surface; $|\vec{q}\,|=k_{\rm F}$ ($\eta=0$),
we take the limit $M_{\rm B} \rightarrow \infty$ ($k_{\rm F}/M_{\rm B} \rightarrow 0$) in the second order scattering amplitude, and obtain the result
\begin{eqnarray}
&& \lim_{M_{\rm B} \rightarrow \infty} \lim_{\eta \rightarrow 0} \left( -i{\cal M}^{(2)}_{\rm B} \right)_{\rm{rel}} \\
&=& 
 \left( -iG_{\rm B}M_{\rm B} \right)^2 \delta_{a'a}\delta_{b'b} i \frac{1}{2M_{\rm B}} \frac{k_{\rm F}^2}{4\pi^2} 4 \left( 1-\frac{1}{n^2} \right)
 \bar{u}_{q,a'} v\hspace{-0.5em}/ (-4\Lambda) \gamma^{0} v\hspace{-0.5em}/ u_{q,a}
    \nonumber \\
&& + \left( -iG_{\rm B}M_{\rm B} \right)^2 (\vec{\lambda}_{\rm f})_{a'a} \cdot (\vec{\lambda}_{\rm B})_{b'b} \nonumber \\
&& \times i \frac{1}{2M_{\rm B}} \frac{k_{\rm F}^2}{4\pi^2} \bar{u}_{q,a'} v\hspace{-0.5em}/ \left[ \left\{ -4\left(n-\frac{4}{n}\right) + n\Lambda^{2} - 2n \log \frac{k_{\rm F}}{M_{\rm B}} + 2n \log \Lambda + 2n(1-\log 2)\right\}\gamma^{0} - n \hat{q}\cdot\vec{\gamma} \right] \nonumber \\
&& \times v\hspace{-0.5em}/ u_{q,a}, \nonumber
\end{eqnarray}
with the relativistic spinor $u_{q}$. 
As in the case of the nonrelativistic fermions, we find again that the factor $\log M_{\rm B}$ exists.
The second order scattering amplitude contains $G_{\rm B}^{\,2}M_{\rm B} \log M_{\rm B}$ for large $M_{\rm B}$, while the first order one contains $G_{\rm B} M_{\rm B}$.
Therefore, the second order scattering amplitude can be larger than the first order one in the limit of large $M_{\rm B}$ and the perturbation is not applicable for any small coupling constant $G_{\rm B}$ in this limit.
The critical mass of $M_{\rm B}$ for which the perturbation cannot be applied is given by
\begin{eqnarray}
 M_{\rm B}^{\rm{cr,rel}} = k_{\rm F} \exp \left( \frac{1}{n |G_{\rm B}|} \frac{4\pi^{2}}{k_{\rm F}^{2}} \right).
 \label{eq:critical_mass_boson_rela}
\end{eqnarray}
We keep in mind that this number should be not be considered seriously as mentioned below Eq.~(\ref{eq:critical_mass_boson_nonrela}).

\section{Heavy impurity fermion} \label{sec:fermion}

We consider a charm (bottom) quark in quark matter where up, down and strangeness quarks compose the Fermi surface.
Although the quark matter is much different from the nuclear matter, we will find again that the scattering amplitude in large limit of the heavy quark mass has a behavior similar to that obtained in the previous section.
As an interaction between the charm (bottom) quark and the light quark, we suppose the vector current interaction with a color factor of $\vec{\lambda}_{\rm f} \cdot \vec{\lambda}_{\rm F}$ (see below) with SU(3) color symmetry.
Assuming a small coupling constant, we analyze the scattering amplitude up to one-loop level.
As in the previous section, we extend SU(3) symmetry to SU($n$) symmetry with an arbitrary integer $n \ge 2$.
We use the common notations as in the previous section, except for the the coupling constant and the mass of the heavy impurity fermion.

\subsection{Interaction with SU($n$) symmetry}

We consider the interaction Lagrangian for the fermion composing the Fermi surface and the heavy impurity fermion
\begin{eqnarray}
 {\cal L_{\rm{F, int}}} =
  - \frac{G_{\rm F}}{2} \sum_{j=1}^{n^2-1} ( \bar{\psi} \gamma_{\mu} \lambda_{\rm{f}}^{j} \psi )
   (\bar{\Psi} \gamma^{\mu} \lambda_{\rm{F}}^{j} \Psi)
\label{eq:int_fermion}
\end{eqnarray}
where the fermion field $\psi=(\psi_{1}\,\cdots,\psi_{n})^{\rm t}$ and the heavy impurity fermion field $\Psi=(\Psi_{1},\cdots,\Psi_{n})^{\rm t}$ belong to the fundamental representation of SU($n$) symmetry.
The $n \times n$ matrices $\lambda_{\rm{f}}^{j}$ and $\lambda_{\rm{F}}^{j}$ ($j=1,\cdots,n^2-1$) are the generators of SU($n$) group. 
In this section, we use $G_{\rm F}/2$ as the coupling constant.
For the heavy impurity fermion, it is convenient to separate the momentum of the heavy impurity fermion as $P=M_{\rm F}v+k$ with the heavy impurity fermion mass $M_{\rm F}$, the four-dimensional velocity $v$ with a condition $v^2=1$, and the residual momentum $k$.
This convention will be used later.

\subsection{Scattering amplitude}

Let us consider the scattering amplitude of the heavy impurity fermion and the fermion composing the Fermi surface.
The diagrams are the same as in Figs.~\ref{fig:Fig1} and \ref{fig:Fig2}, provided that the heavy impurity boson should be read as the heavy impurity fermion.

The first oder scattering amplitude (the Born term) is given as
\begin{eqnarray}
-i {\cal M}^{(1)}_{\rm F} = -i\frac{G_{\rm F}}{2} \bar{u}_{q,a'} \gamma_{\mu} (\vec{\lambda}_{\rm{f}})_{a'a} u_{q,a} \bar{u}_{P,b'} \gamma^{\mu} (\vec{\lambda}_{\rm{F}})_{b'b} u_{P,b},
\end{eqnarray}
where $u_{q}$ ($\bar{u}_{q}$) is the spinor wave function for the initial (final) fermion with four-dimensional momentum $q$, and $u_{P}$ ($\bar{u}_{P}$) is the spinor wave function for the initial (final) heavy impurity fermion with four-dimensional momentum $P$.
As discussed in the previous section, it is reasonable to suppose that the initial and final states have the same momentum.

The second oder scattering amplitude at one-loop level is given as
\begin{eqnarray}
-i{\cal M}^{(2)}_{\rm F}=-i{\cal M}^{(2)}_{\rm F} ({\rm{Fig.~\ref{fig:Fig2} (a)}})-i{\cal M}^{(2)}_{\rm F} ({\rm{Fig.~\ref{fig:Fig2} (b)}}),
\end{eqnarray}
with
\begin{eqnarray}
&& -i {\cal M}^{(2)}_{\rm F} ({\rm{Fig.~\ref{fig:Fig2} (a)}}) \nonumber \\
&=& \left(-i\frac{G_{F}}{2}\right)^2  \left\{ 4\left(1-\frac{1}{n^2}\right) \delta_{a'a} \delta_{b'b} +\left( -\frac{4}{n} \right) (\vec{\lambda}_{\rm f})_{a'a} \! \cdot \! (\vec{\lambda}_{\rm F})_{b'b} \right\} \nonumber \\
&& \times \bar{u}_{q,a'} \gamma_{\mu} \int \frac{{\rm d}^{4}k}{(2\pi)^4} (k\hspace{-0.5em}/+m)
\left[ \frac{i}{k^{2}-m^{2}+i\epsilon} - 2\pi \delta(k^2-m^2) \theta(k_0) \theta(k_{\rm F}-|\vec{k}\,|) \right] \gamma_{\nu} u_{q,a} \nonumber \\
&& \times \bar{u}_{P,b'} \gamma^{\mu} \frac{i}{(P+q-k)^2-M_{\rm F}^{2}+i\epsilon} (P\hspace{-0.5em}/+q\hspace{-0.5em}/-k\hspace{-0.5em}/+M_{\rm F}) \gamma^{\nu} u_{P,b}
\end{eqnarray}
in Fig.~\ref{fig:Fig2} (a), and 
\begin{eqnarray}
&& -i {\cal M}^{(2)}_{\rm F} ({\rm{Fig.~\ref{fig:Fig2} (b)}}) \nonumber \\
&=& \left( -i\frac{G_{\rm F}}{2} \right)^2 \left\{ 4\left(1-\frac{1}{n^2}\right) \delta_{a'a} \delta_{b'b} + 2 \left( n-\frac{2}{n} \right) (\vec{\lambda}_{\rm f})_{a'a} \! \cdot \! (\vec{\lambda}_{\rm F})_{b'b} \right\} \nonumber \\
&& \times \bar{u}_{q,a'} \gamma_{\mu} \int \frac{{\rm d}^{4}k}{(2\pi)^4} (k\hspace{-0.5em}/+m)
\left[ \frac{i}{k^{2}-m^{2}+i\epsilon} - 2\pi \delta(k^2-m^2) \theta(k_0) \theta(k_{\rm F}-|\vec{k}\,|) \right] \gamma_{\nu} u_{q,a} \nonumber \\
&& \times \bar{u}_{P,b'} \gamma^{\nu} \frac{i}{(P+q-k)^2-M_{\rm F}^{2}+i\epsilon} (P\hspace{-0.5em}/+q\hspace{-0.5em}/-k\hspace{-0.5em}/+M_{\rm F}) \gamma^{\mu} u_{P,b}
\end{eqnarray}
in Fig.~\ref{fig:Fig2} (b).
In deriving the above equations, we use the identities in Eqs.~(\ref{eq:identity1}) and (\ref{eq:identity2}).
Because the heavy impurity fermion is sufficiently massive, the matrices $\gamma^{\rho}$ at vertices acting for the field $\Psi$ is replaced to $v^{\rho}$.
Performing the integrals for $k_{0}$, we obtain the result for the finite density part
\begin{eqnarray}
&& -i{\cal M_{\rm F}}^{(2)} \nonumber \\
&=&  \left( -i\frac{G_{\rm F}}{2} \right)^2 4 \left( 1-\frac{1}{n^2} \right) \delta_{a'a} \delta_{b'b} \label{eq:amp2_fermion} \\
&& \times 2M_{\rm F} i \left[ \int_{|\vec{k}|\ge k_{\rm F}} \frac{{\rm d}^{3} \vec{k}}{(2\pi)^{3}} \frac{1}{2\epsilon_{k}} \frac{1}{(P_{0}+q_{0}-\epsilon_{k})^2 - E_{P+q-k}^{2} + i \varepsilon} \bar{u}_{q,a'} v\hspace{-0.5em}/ \left.(k\hspace{-0.5em}/+m)\right|_{k_{0}=\epsilon_{k}} v\hspace{-0.5em}/ u_{q,a} \right. \nonumber \\
&& \left. - \int_{|\vec{k}|\le k_{\rm F}} \frac{{\rm d}^{3} \vec{k}}{(2\pi)^{3}} \frac{1}{2\epsilon_{k}} \frac{1}{(P_{0}-q_{0}+\epsilon_{k})^2 - E_{P-q+k}^{2} + i \varepsilon} \bar{u}_{q,a'} v\hspace{-0.5em}/ \left.(k\hspace{-0.5em}/+m)\right|_{k_{0}=\epsilon_{k}} v\hspace{-0.5em}/ u_{q,a} \right]  \nonumber \\
&& \times \bar{u}_{P,b'} \frac{1+v\hspace{-0.5em}/}{2} u_{P,b} \nonumber \\
&& + \left( -i\frac{G_{\rm F}}{2} \right)^2 (\vec{\lambda}_{\rm f})_{a'a} \cdot (\vec{\lambda}_{\rm F})_{b'b} \nonumber \\
&& \times 2M_{\rm F} i \left[ \left( -\frac{4}{n} \right) \int_{|\vec{k}|\ge k_{\rm F}} \frac{{\rm d}^{3} \vec{k}}{(2\pi)^{3}} \frac{1}{2\epsilon_{k}} \frac{1}{(P_{0}+q_{0}-\epsilon_{k})^2 - E_{P+q-k}^{2} + i \varepsilon} \bar{u}_{q,a'} v\hspace{-0.5em}/ \left.(k\hspace{-0.5em}/+m)\right|_{k_{0}=\epsilon_{k}} v\hspace{-0.5em}/ u_{q,a} \right. \nonumber \\
&& \left. - 2\left(n-\frac{2}{n}\right) \int_{|\vec{k}|\le k_{\rm F}} \frac{{\rm d}^{3} \vec{k}}{(2\pi)^{3}} \frac{1}{2\epsilon_{k}} \frac{1}{(P_{0}-q_{0}+\epsilon_{k})^2 - E_{P-q+k}^{2} + i \varepsilon} \bar{u}_{q,a'} v\hspace{-0.5em}/ \left.(k\hspace{-0.5em}/+m)\right|_{k_{0}=\epsilon_{k}} v\hspace{-0.5em}/ u_{q,a} \right] \nonumber \\
&& \times \bar{u}_{P,b'} \frac{1+v\hspace{-0.5em}/}{2} u_{P,b}, \nonumber
\end{eqnarray}
where the terms suppressed in the limit of large $M_{\rm F}$ are dropped.
Here we define $E_{P}=\sqrt{\vec{P}^2+M_{\rm F}^{\,2}}$ as the energy of the heavy impurity fermion.
We use the condition that the initial and final heavy impurity fermion is the on-mass-shell state which is resting in the medium; $P=M_{\rm F} v$ with $v^{\mu}=(1,\vec{0})$.
Then, the scattering amplitude up to one-loop level is given by Eq.~(\ref{eq:amp_tot}) with changing the subscript from ${\rm B}$ to ${\rm F}$.

As for the heavy fermion close to the on-mass-shell state, we know that the propagator is given as $(P\hspace{-0.5em}/-M_{\rm F})^{-1} \simeq (v\hspace{-0.5em}/+1) (2 v\cdot k)^{-1}$ from the decomposition of the momentum $P=M_{\rm F}v+k$.
Then, we may expect that the second order scattering amplitude could be proportional to $G_{\rm F}^2 ({\rm{vertex}}) \times M_{\rm F}^{0} ({\rm{propagator}}) =G_{\rm F}^{2}$ when the momentum cutoff in the loop is fixed, while the first order one is proportional to $G_{\rm F}$.
Hence, it seems that the perturbation might be applicable for the small coupling constant $G_{\rm F}$.
However, as in the previous section, we will find that there exists a logarithmic factor in the second order scattering amplitude and that the perturbation for any small coupling constant $G_{\rm F}$ is not applicable in the heavy mass limit $M_{\rm F} \rightarrow \infty$.

\subsection{Large mass limit of heavy impurity fermion}

We consider the large mass limit for the heavy impurity fermion ($M_{\rm F} \rightarrow \infty$).
To analyze the second order scattering amplitude,
we consider the two cases for the fermions composing the Fermi surface; nonrelativistic and relativistic fermions.
Although the relativistic fermions are much likely in quark matter, the nonrelativistic fermions will be found to be useful.

\subsubsection{Nonrelativistic fermions}

We consider that the fermion mass $m$ is sufficiently large than the Fermi momentum $k_{\rm F}$,
and use $u_{q} \simeq \sqrt{2m} (\chi,0)^{\rm t}$ as nonrelativistic spinor.
We suppose that the fermion lies on the Fermi surface ($|\vec{q}\,|=k_{\rm F}$; $\eta=0$). 
We note that the condition $m / M_{\rm F} \ll 1$ is still kept.
By considering the large mass limit of the heavy impurity fermion,
we obtain the result
\begin{eqnarray}
&& \lim_{M_{\rm B} \rightarrow \infty} \lim_{\eta \rightarrow 0} \left( -i{\cal M}^{(2)}_{\rm B} \right)_{\rm{nonrel}}
\label{eq:amp2_fermion_non-rela} \\
&=&  \left( -i\frac{G_{\rm F}}{2} \right)^2  \delta_{a'a} \delta_{b'b} \nonumber \\ 
&& \times i  \frac{2mk_{\rm F}}{4\pi^{2}} 4 \left( 1-\frac{1}{n^2} \right) \left[ -4\Lambda + \log \left( 1+\frac{\Lambda}{2} \right) - \log \left( 1-\frac{\Lambda}{2} \right) \right] \bar{u}_{q,a'} \frac{1+v\hspace{-0.5em}/}{2} u_{q,a} \nonumber \\
&& \times \bar{u}_{P,b'} \frac{1+v\hspace{-0.5em}/}{2} u_{P,b} \nonumber \\
&& + \left(-i \frac{G_{\rm F}}{2} \right)^2 (\vec{\lambda}_{\rm f})_{a'a} \cdot (\vec{\lambda}_{\rm F})_{b'b} \nonumber \\
&& \times i \frac{2mk_{\rm F}}{4\pi^{2}} \left[ -4\left( n-\frac{4}{n} \right) \Lambda + 2n \log \Lambda + \left(-\frac{4}{n}\right) \log \left( 1+\frac{\Lambda}{2} \right) - 2 \left( n-\frac{2}{n} \right) \log \left( 1-\frac{\Lambda}{2} \right) \right. \nonumber \\
&& \left.- 2n \log \frac{m}{M_{\rm F}} +2n (1-\log2 )\right] \bar{u}_{q,a'} \frac{1+v\hspace{-0.5em}/}{2} u_{q,a} \nonumber \\
&& \times \bar{u}_{P,b'} \frac{1+v\hspace{-0.5em}/}{2} u_{P,b}. \nonumber
\end{eqnarray}
We find the logarithmic factor $\log M_{\rm F}$ in the term proportional to $(\vec{\lambda}_{\rm f})_{a'a} \cdot (\vec{\lambda}_{\rm F})_{b'b}$.
The first oder scattering amplitude is proportional to $G_{\rm F}$, while the second oder scattering amplitude contains $G_{\rm F}^2 \log M_{\rm F}$ in the limit of large $M_{\rm F}$.
It means that, due to the presence of $\log M_{\rm F}$, the second oder scattering amplitude becomes larger than the first order one in the limit of large $M_{\rm F}$.
Therefore, the perturbation for any small coupling $G_{\rm F}$ fails in the system in this limit.
The critical mass of $M_{\rm F}$ is given in Eq.~(\ref{eq:critical_mass_boson_nonrela}) with a replacement from $G_{\rm B}$ to $G_{\rm F}$ and from $M_{\rm B}$ to $M_{\rm F}$.

The result of the heavy impurity fermion is analogous to that of the heavy impurity boson.
As discussed in the previous section, the reason why the singularity from the heavy mass arises is given by two reasons.
First, the existence of the factor $\vec{\lambda}_{\rm f} \cdot \vec{\lambda}_{\rm F}$ in the interaction Lagrangian Eq.~(\ref{eq:int_fermion}) is important.
Indeed, we see that there is no logarithmic behavior in the term with $\delta_{a'a}\delta_{b'b}$ in Eq.~(\ref{eq:amp2_fermion_non-rela}).
Second, the large mass of the heavy impurity fermion is also important.
Indeed, when the mass of the heavy impurity fermion happens to be equal to the mass of the fermion composing the Fermi surface, $M_{\rm F}=m$,
we obtain from Eq.~(\ref{eq:amp2_fermion})
\begin{eqnarray}
&& \lim_{M_{\rm F} \rightarrow m} \lim_{\eta \rightarrow 0} \left(-i{\cal M}^{(2)}_{\rm F}\right)_{\rm{nonrel}} 
\label{eq:amp2_fermion_non-rela_2} \\
&=&  \left(-i \frac{G_{\rm F}}{2} \right)^2  \delta_{a'a} \delta_{b'b} \nonumber \\ 
&& \times i  \frac{k_{\rm F} \Lambda}{4\pi^{2}} 4 \left( 1-\frac{1}{n^2} \right) \left[ -\left(1+\frac{\Lambda}{2}\right) \frac{1}{2} \log \frac{2+\Lambda}{\Lambda} + \left( 1-\frac{\Lambda}{2} \right) \frac{1}{2} \log \frac{2-\Lambda}{\Lambda} \right] \bar{u}_{q,a'} \frac{1+v\hspace{-0.5em}/}{2} u_{q,a} \nonumber \\
&& \times \bar{u}_{P,b'} \frac{1+v\hspace{-0.5em}/}{2} u_{P,b} \nonumber \\
&& + \left( -i\frac{G_{\rm F}}{2} \right)^2 (\vec{\lambda}_{\rm f})_{a'a} \cdot (\vec{\lambda}_{\rm F})_{b'b} \nonumber \\
&& \times i \frac{k_{\rm F}\Lambda}{4\pi^{2}} \left[ n - \left(-\frac{4}{n} \right) \left( 1+ \frac{\Lambda}{2} \right) \frac{1}{2} \log \frac{2+\Lambda}{\Lambda} + 2\left( n-\frac{2}{n} \right) 2 \left( 1-\frac{\Lambda}{2} \right) \frac{1}{2} \log\frac{2-\Lambda}{\Lambda} \right] \bar{u}_{q,a'} \frac{1+v\hspace{-0.5em}/}{2} u_{q,a} \nonumber \\
&& \times \bar{u}_{P,b'} \frac{1+v\hspace{-0.5em}/}{2} u_{P,b}. \nonumber
\end{eqnarray}
We confirm that there is no singular term for $M_{\rm F}=m$.

\subsubsection{Relativistic fermions}

We consider that the fermions composing the Fermi surface are relativistic.
Supposing the massless fermions, we replace the nonrelativistic spinor to the relativistic spinor $u_{q} = \sqrt{E_{q}} (\chi, \vec{\sigma} \! \cdot \! \hat{q} \, \chi  )^{\rm t}$ with $E_{q} = |\vec{q}\,|$ and $\hat{q}=\vec{q}/|\vec{q}\,|$.
We use the expansion parameter $k_{\rm F}/M_{\rm F}$ in the limit of the large mass of the heavy impurity fermion.
The resulting form of the second oder scattering amplitude is
\begin{eqnarray}
&& \lim_{M_{\rm F} \rightarrow \infty} \lim_{\eta \rightarrow 0} \left( -i{\cal M}^{(2)}_{\rm F} \right)_{\rm{rel}} 
\label{eq:amp2_fermion_rela} \\
&=& 
 \left(-i\frac{G_{\rm F}}{2}\right)^2 \delta_{a'a}\delta_{b'b} i  \frac{k_{\rm F}^2}{4\pi^2} 4 \left( 1-\frac{1}{n^2} \right) 
 \bar{u}_{q,a'} v\hspace{-0.5em}/ (-4\Lambda) \gamma^{0} v\hspace{-0.5em}/ u_{q,a}
   \bar{u}_{P,b'} \frac{1+v\hspace{-0.5em}/}{2} u_{P,b} \nonumber \\
&& + \left(-i\frac{G_{\rm F}}{2}\right)^2 (\vec{\lambda}_{\rm f})_{a'a} \cdot (\vec{\lambda}_{\rm F})_{b'b} \nonumber \\
&& \times i \frac{k_{\rm F}^2}{4\pi^2} \bar{u}_{q,a'} v\hspace{-0.5em}/ \left[ \left\{ -4\left(n-\frac{4}{n}\right) + n\Lambda^{2} - 2n \log \frac{k_{\rm F}}{M_{\rm F}} + 2n \log \Lambda + 2n(1-\log 2)\right\}\gamma^{0} - n \hat{q}\cdot\vec{\gamma} \right] v\hspace{-0.5em}/ u_{q,a} \nonumber \\
&& \times \bar{u}_{P,b'} \frac{1+v\hspace{-0.5em}/}{2} u_{P,b}. \nonumber
\end{eqnarray}
We again obtain the factor $\log M_{\rm F}$. 
Therefore, 
 the perturbation breaks down in the heavy mass limit.
The critical mass of $M_{\rm F}$ is given in Eq.~(\ref{eq:critical_mass_boson_rela}) with a replacement from $G_{\rm B}$ to $G_{\rm F}$ and from $M_{\rm B}$ to $M_{\rm F}$.

\section{Discussion} \label{sec:discussion}

We have shown that
 the perturbation breaks down for any small coupling constant due to the logarithmic enhancement in the limit of large mass of the heavy impurity particles for both boson and fermion.
We have obtained those results for the case that the scattering 
fermion is in the ground state, namely $|\vec{q}\,|=k_{\rm F}$ ($\eta=0$), 
at zero temperature.
For the case that the scattering fermion is in the excited state with $|\vec
{q}\,|>k_{\rm F}$ ($\eta>0$), the result is modified in a qualitative 
manner.
In this case, 
there is no logarithmic term of the mass of the heavy impurity particles.
In the limit of small $\eta$ ($\eta \rightarrow 0$), instead, it gives a new 
logarithmic term, $\log \eta$, as shown explicitly in Appendix~\ref{sec:Kondo}.
Therefore, 
due to the logarithmic enhancement by the energy of the scattering fermions,
 the system becomes a strongly interacting one.
Indeed, this is the Kondo problem as mentioned in the introduction \cite{Kondo:1964,Hewson}.
In both approaches in the different two limits, the reason for the presence of logarithmic terms will be found in the dynamics of the particles and the holes in the loop contribution.
However, we do not pursue the problem in the present discussion, because we must go beyond the one-loop level for more detailed analysis.
Instead, we shortly discuss the possible phenomena in nuclear and quark matter including charm and bottom flavor.

Let us consider the $\bar{D}$ and $B$ mesons embedded in nuclear matter at zero temperature.
When we apply the heavy mass limit to this system, we are inevitably faced with the strong coupling problem as we have shown in section \ref{sec:boson}.
We remember that the second order scattering amplitude Eq.~(\ref{eq:amp2_boson_non-rela}), namely the second order vertex, has a logarithmic enhancement by $M_{\rm B}$ in the limit of large $M_{\rm B}$.
The logarithmic behavior appears only in the term with the isospin factor $\vec{\lambda}_{\rm f} \cdot \vec{\lambda}_{\rm B}$, and does not in the term without the isospin factor.
Therefore, we expect that the isospin-dependent interaction is much enhanced than the isospin-independent one. 
The strong isospin dependence may cause some change of the structure of nuclear matter, because the properties of nuclear matter is much sensitive to the isospin symmetry.
We note that, however, the mass modifications of $\bar{D}$ and $B$ mesons in nuclear medium are not suffered regardless to the logarithmic enhancement in the scattering amplitude.
This is seen by the fact that, when the fermion outer lines are closed in Fig.~\ref{fig:Fig2}, the contribution from the term proportional to the factor $\vec{\lambda}_{\rm f} \cdot \vec{\lambda}_{\rm B}$ becomes zero.

As explained in the introduction, there have been discussions about charmed (bottom) nuclei where charm (bottom) hadrons, such $\Lambda_{\rm c}$ and $\Sigma_{\rm c}$ ($\Lambda_{\rm b}$ and $\Sigma_{\rm b}$) baryons, $\bar{D}$ and $D$ ($B$ and $\bar{B}$) mesons, are bound in atomic nuclei.
We have to note that the current formalism
is not applied, unless the impurity particle belongs to the fundamental representation of SU($n$) symmetry ($n \ge 2$).
For example, $\Lambda_{\rm c}$ ($\Lambda_{\rm b}$) baryon is an isospin singlet state and $\Sigma_{\rm c}$ ($\Sigma_{\rm b}$) baryon is an isospin triplet state, 
and hence they cannot be applied. 
As candidates, we may consider $D$ ($\bar{B}$) meson in nuclear matter, because it is an isospin doublet state.
However, it is actually unstable in nuclear matter, because it can decay through the transitions by two-body processes $DN  \rightarrow \pi \Sigma_{\rm c}^{(\ast)} $ ($\bar{B}N  \rightarrow \pi \Sigma_{\rm b}^{(\ast)} $) as well as three-body absorption processes $DNN  \rightarrow \Lambda_{\rm c}^{(\ast)}N$, $\Sigma_{\rm c}^{(\ast)}N $ ($\bar{B}NN  \rightarrow \Lambda_{\rm b}^{(\ast)}N$, $\Sigma_{\rm b}^{(\ast)}N $) which are opened below the thresholds.
We will need to extend the formalism to include such complex processes.
In contrast, $\bar{D}$ and $B$ mesons in nuclear matter have no open channel below the thresholds, because there is no annihilation channel for light quark and antiquark pairs.
As a result, we conclude that $\bar{D}$ and $B$ mesons
 are unique hadrons for which the present discussion can be applied.

Similar discussion will be applied to the charm and bottom quarks in quark matter at zero temperature as presented in section \ref{sec:fermion}.
We remind us that, in general, the coupling in the interaction between quarks becomes smaller at high density limit, due to the asymptotic freedom of QCD.
According to the present discussion, however, charm and bottom quarks can interact strongly with the light quarks composing the Fermi surface, because there is a logarithmic enhancement by charm (bottom) quark mass in the second order scattering amplitude.
Hence we will need to consider the strong coupling problem for the quark matter including charm and bottom flavor as far as the heavy quark mass limit is adopted.

In both cases of $\bar{D}$ and $B$ mesons in nuclear matter and charm and bottom quarks in quark matter, we find that the scattering amplitudes have a common property; the logarithmic enhancement by the mass of the heavy impurity particle.
It will be an interesting future problem to study the strong coupling effect by logarithmic enhancement in the dynamics of nuclear matter and quark matter with charm and bottom flavor.

\section{Summary}

We discuss the dynamics of the heavy impurity particle (hadron or quark) embedded in finite density medium at zero temperature.
We suppose that the fermions composing the Fermi surface and the heavy impurity particle belong to the fundamental representation of SU($n$) symmetry ($n \ge 2$), and that they interact through the vector current interaction with a small coupling constant.
As systems, we consider a $\bar{D}$ ($B$) meson in nuclear matter with isospin symmetry ($n=2$), and a charm (bottom) quark in quark matter with color symmetry ($n=3$).
We calculate the scattering amplitude for the fermion and the embedded heavy impurity particle.
We analyze the large mass limit of the heavy impurity particle, and find that the second order scattering amplitude at one-loop level contains a logarithmic term of the mass of the heavy impurity particle.
Due to the logarithmic enhancement in the heavy mass limit, the perturbation breaks down for any small coupling constant and the system becomes a strongly interacting one. 

When the present result is applied to $\bar{D}$ ($B$) meson in nuclear matter and charm (bottom) quark in quark matter,
we expect that they are strongly interacting systems as far as the heavy mass limit is concerned. 
To study more details, for example,  
we will need to consider more realistic interaction such as the long range pion exchange potential for $\bar{D}$ ($B$) meson and the gluon exchange for charm (bottom) quark.
It will be interesting to study such problems for future experiments in high energy accelerator facilities with high momentum hadron beam in such as J-PARC, GSI-FAIR \cite{Golubeva:2002au} as well as in relativistic heavy ion collisions in RHIC and LHC \cite{Cho:2010db,Cho:2011ew} and so on.

\begin{acknowledgments}
This work is supported in part by Grant-in-Aid for Scientific Research on 
Priority Areas ``Elucidation of New Hadrons with a Variety of Flavors 
(E01: 21105006)" (S.Y.) and by ``Grant-in-Aid for Young Scientists (B)
22740174" (K.S.), from 
the ministry of Education, Culture, Sports, Science and Technology of
Japan.
\end{acknowledgments}

\begin{appendix}

\section{A calculation for momentum integrals}
\label{sec:calc}

We give a useful equation to derive Eq.~(\ref{eq:amp2_boson_non-rela}) from Eq.~(\ref{eq:amp2_boson}).
The three-dimensional $\vec{k}$-integrals in Eq.~(\ref{eq:amp2_boson}) is reduced to two dimensional integrals with the radial component $k \equiv |\vec{k}\,|$ and the angular component $t \equiv \vec{q} \!\cdot\! \vec{k}/|\vec{q}\,| |\vec{k}\,|$.
When we expand up to the order of $m/M$ for small $m/M \ll1$ (at least $m/M<1/2$ is assumed) and neglect the terms with higher order ${\cal O}\left((m/M)^2\right)$, we obtain 
\begin{eqnarray}
&& \int_{|\vec{k}\,| \ge k_{\rm F}} \frac{{\rm d}^{3}\vec{k}}{(2\pi)^3}
\frac{1}{\frac{1}{2m}\vec{q}\,\,^2-\frac{1}{2m}\vec{k}\,^2-\frac{1}{2M}|\vec{q}-\vec{k}|^2 + i \varepsilon} \nonumber \\
&=& \frac{1}{4\pi^2} \int_{-1}^{1} {\rm d}t \int_{k_{\rm F}}^{\Lambda_{\rm{high}}} {\rm d} k \,k^2
\frac{1}{\frac{1}{2m} q\,^2-\frac{1}{2m} k\,^2-\frac{1}{2M}(q^2+k^2-2qkt) + i \varepsilon} \nonumber \\
&=&
 \frac{1}{4\pi^2} \left( - \frac{2m}{1+m/M} \right) \nonumber \\
&& \times \int_{-1}^{1} {\rm d}t \left[ (\Lambda_{\rm{high}}-k_{\rm F}) 
+ \frac{q}{2} \left( 1+\frac{m}{M} (2t-1) \right)  \log \frac{\Lambda_{\rm{high}}-q\left( 1- (1-t)m/M \right)}{\left| q\left( 1-(1-t)m/M \right)-k_{\rm F} \right|} \right. \nonumber \\
&& \hspace{8.3em} \left. - \frac{q}{2} \left( 1-\frac{m}{M} (2t+1) \right)  \log \frac{\Lambda_{\rm{high}}+q\left( 1- (1+t)m/M \right)}{k_{\rm F}+q\left( 1-(1+t)m/M \right)} \right] \nonumber \\
&& -i \pi  \frac{1}{4\pi^2} \frac{2m}{1+m/M} \left\{
\begin{array}{l}
 \int_{-1}^{1} {\rm d}t \frac{q}{2} \left( 1+ \frac{m}{M}(2t-1) \right) \hspace{2.8em} {\rm for} \hspace{1em} q(1-2m/M)>k_{\rm F} \\
 \int_{t_{\rm F}}^{1} {\rm d}t \frac{q}{2} \left( 1+ \frac{m}{M}(2t-1) \right) \hspace{3em} {\rm for} \hspace{1em} q(1-2m/M)<k_{\rm F}
\end{array}
\right. \nonumber \\
&& + {\cal O}\left((m/M)^2\right),
\end{eqnarray}
for $|\vec{k}\,|>k_{\rm F}$ with higher cutoff parameter $\Lambda_{\rm{high}} > k_{\rm F}$, and
\begin{eqnarray}
&& \int_{|\vec{k}\,| \le k_{\rm F}} \frac{{\rm d}^{3}\vec{k}}{(2\pi)^3}
\frac{1}{\frac{1}{2m}\vec{q}\,\,^2-\frac{1}{2m}\vec{k}\,^2+\frac{1}{2M}|\vec{q}-\vec{k}|^2 - i \varepsilon} \nonumber \\
&=& \frac{1}{4\pi^2} \int_{-1}^{1} {\rm d}t \int_{\Lambda_{\rm{low}}}^{k_{\rm F}} {\rm d} k \,k^2
\frac{1}{\frac{1}{2m} q\,^2-\frac{1}{2m} k\,^2+\frac{1}{2M}(q^2+k^2-2qkt) - i \varepsilon} \nonumber \\
&=&
 \frac{1}{4\pi^2} \left( - \frac{2m}{1+m/M} \right) \nonumber \\
&& \times \int_{-1}^{1} {\rm d}t \left[ (k_{\rm F}-\Lambda_{\rm{low}}) 
+ \frac{q}{2} \left( 1-\frac{m}{M} (2t-1) \right)  \log \frac{ q\left( 1+(1-t)m/M \right)-k_{\rm F} }{q\left( 1+ (1-t)m/M \right) -\Lambda_{\rm{low}}} \right. \nonumber \\
&& \hspace{8.3em} \left. - \frac{q}{2} \left( 1+\frac{m}{M} (2t+1) \right)  \log \frac{k_{\rm F}+q\left( 1+(1+t)m/M \right)}{\Lambda_{\rm{low}}+q\left( 1+ (1+t)m/M \right)} \right] \nonumber \\
&& + {\cal O}\left((m/M)^2\right),
\end{eqnarray}
for $|\vec{k}\,|>k_{\rm F}$ with lower cutoff parameter $\Lambda_{\rm{low}} < k_{\rm F}$.
Here we define $q=|\vec{q}\,|$ and
\begin{eqnarray}
t_{\rm F} = - \frac{1}{2} \left( \frac{M}{m} -1 \right) \frac{q}{k_{\rm F}} + \frac{1}{2} \left( \frac{M}{m} + 1 \right) \frac{k_{\rm F}}{q}.
\end{eqnarray}

\section{The Kondo problem revisited}
\label{sec:Kondo}


For the heavy impurity boson, in the text, we consider that the scattering fermion is in the ground state with the condition $|\vec{q}\,|=k_{\rm F}$ ($\eta=0$).
Let us discuss the case that the fermion is not the ground state but is in the excited state which energy lies above the Fermi surface; $|\vec{q}\,| > k_{\rm F}$ ($\eta>0$).
We expand the second order scattering amplitude $-i{\cal M}^{(2)}_{\rm B}$ in Eq.~(\ref{eq:amp2_boson}) by large $M_{\rm B}$ with keeping $\eta$ fixed, and take the limit for  small $\eta$.
For nonrelativistic fermions, leaving only the leading terms, we obtain
\begin{eqnarray}
&& \lim_{\eta \rightarrow 0} \lim_{M_{\rm B} \rightarrow \infty} \left( -i{\cal M}^{(2)}_{\rm B} \right)_{\rm{nonrel}} 
\label{eq:amp2_boson_non-rela_3} \\
&=&  \left( -iG_{\rm B}M_{\rm B} \right)^2  \delta_{a'a} \delta_{b'b} \nonumber \\ 
&& \times i 4 \left( 1-\frac{1}{n^2} \right) \frac{1}{2M_{\rm B}} \frac{2mk_{\rm F}}{4\pi^{2}} \left[ -4\Lambda + \log \left( 1+\frac{\Lambda}{2} \right) - \log \left( 1-\frac{\Lambda}{2} \right) - i\pi \right] \bar{u}_{q,a'} \frac{1+v\hspace{-0.5em}/}{2} u_{q,a} \nonumber \\
&& + \left( -iG_{\rm B}M_{\rm B} \right)^2 (\vec{\lambda}_{\rm f})_{a'a} \cdot (\vec{\lambda}_{\rm B})_{b'b} \nonumber \\
&& \times i \frac{1}{2M_{\rm B}} \frac{2mk_{\rm F}}{4\pi^{2}} \left[ -4\left( n-\frac{4}{n} \right) \Lambda + 2n \log \Lambda + \left(-\frac{4}{n}\right) \log \left( 1+\frac{\Lambda}{2} \right) - 2 \left( n-\frac{2}{n} \right) \log \left( 1-\frac{\Lambda}{2} \right) \right. \nonumber \\
&& \left.- 2n \log \eta - \left(-\frac{4}{n} \right) i \pi \right] \bar{u}_{q,a'} \frac{1+v\hspace{-0.5em}/}{2} u_{q,a}. \nonumber
\end{eqnarray}
Instead of the factor $\log M_{\rm B}$,
there exists the new factor $\log \eta$ in the term proportional to $(\vec{\lambda}_{\rm f})_{a'a} \cdot (\vec{\lambda}_{\rm B})_{b'b}$.
This is a singular term because it prevents a smooth connection from $\eta > 0$ to $\eta = 0$.
Thus, we find again that the perturbation is not applicable for the small coupling constant $G_{\rm B}$ due to the presence of $\log \eta$, so that the system becomes a strongly interacting one.
This is in fact the original Kondo problem
 that the electrons are affected by (pseudo)spin of an impurity atom with infinite mass \cite{Kondo:1964,Hewson}.

As for relativistic fermions, when we consider the excited fermions with $|\vec{q}\,|>k_{\rm F}$ ($\eta>0$), 
 the second order contribution is given by
\begin{eqnarray}
&& \lim_{\eta \rightarrow 0} \lim_{M_{\rm B} \rightarrow \infty} \left( -i{\cal M}^{(2)}_{\rm B} \right)_{\rm{rel}}  \\
&=& 
 \left( -iG_{\rm B}M_{\rm B} \right)^2 \delta_{a'a}\delta_{b'b} i \frac{1}{2M_{\rm B}} \frac{k_{\rm F}^2}{4\pi^2} 4 \left( 1-\frac{1}{n^2} \right)
 \bar{u}_{q,a'} v\hspace{-0.5em}/ (-4\Lambda-i\pi) \gamma^{0} v\hspace{-0.5em}/ u_{q,a}
    \nonumber \\
&& + \left( -iG_{\rm B}M_{\rm B} \right)^2 (\vec{\lambda}_{\rm f})_{a'a} \cdot (\vec{\lambda}_{\rm B})_{b'b} \nonumber \\
&& \times i \frac{1}{2M_{\rm B}} \frac{k_{\rm F}^2}{4\pi^2} \bar{u}_{q,a'} v\hspace{-0.5em}/ \left[ -4\left(n-\frac{4}{n}\right) + n\Lambda^{2} - 2n \log \eta + 2n \log \Lambda + i \pi \frac{4}{n}\right] \gamma^{0} v\hspace{-0.5em}/ u_{q,a}, \nonumber 
\end{eqnarray}
with the relativistic spinor $u_{q}$.
We obtain the factor $\log \eta$ again.
This is the relativistic version of the Kondo problem for the massless fermions.


For the heavy impurity fermion, we discuss in a similar way. 
For the nonrelativistic fermion, we obtain the result
\begin{eqnarray}
&& \lim_{\eta \rightarrow 0} \lim_{M_{\rm F} \rightarrow \infty} \left( -i{\cal M}^{(2)}_{\rm F} \right)_{\rm{nonrel}} \\
&=&  \left(-i\frac{G_{\rm F}}{2}\right)^2  \delta_{a'a} \delta_{b'b} \nonumber \\ 
&& \times i  \frac{2mk_{\rm F}}{4\pi^{2}} 4 \left( 1-\frac{1}{n^2} \right) \left[ -4\Lambda + \log \left( 1+\frac{\Lambda}{2} \right) - \log \left( 1-\frac{\Lambda}{2} \right) - i\pi \right] \bar{u}_{q,a'} \frac{1+v\hspace{-0.5em}/}{2} u_{q,a} \nonumber \\
&& \times \bar{u}_{P,b'} \frac{1+v\hspace{-0.5em}/}{2} u_{P,b} \nonumber \\
&& + \left(-i\frac{G_{\rm F}}{2}\right)^2 (\vec{\lambda}_{\rm f})_{a'a} \cdot (\vec{\lambda}_{\rm F})_{b'b} \nonumber \\
&& \times i \frac{2mk_{\rm F}}{4\pi^{2}} \left[ -4\left( n-\frac{4}{n} \right) \Lambda + 2n \log \Lambda + \left(-\frac{4}{n}\right) \log \left( 1+\frac{\Lambda}{2} \right) - 2 \left( n-\frac{2}{n} \right) \log \left( 1-\frac{\Lambda}{2} \right) \right. \nonumber \\
&& \left.- 2n \log \eta - \left(-\frac{4}{n} \right) i \pi \right] \bar{u}_{q,a'} \frac{1+v\hspace{-0.5em}/}{2} u_{q,a} \nonumber \\
&& \times \bar{u}_{P,b'} \frac{1+v\hspace{-0.5em}/}{2} u_{P,b}. \nonumber
\end{eqnarray}
There exists the factor $\log \eta$ in the terms proportional to $(\vec{\lambda}_{\rm f})_{a'a} \cdot (\vec{\lambda}_{\rm F})_{b'b}$, which gives a logarithmic divergence in the limit of $\eta \rightarrow 0$.
This is again the Kondo problem for the heavy impurity fermion as mentioned in the case of the heavy impurity boson.

When the scattering fermion is relativistic, we set $m=0$ for massless fermions and obtain the result
\begin{eqnarray}
&& \lim_{\eta \rightarrow 0} \lim_{M_{\rm F} \rightarrow \infty} \left( -i{\cal M}^{(2)}_{\rm F} \right)_{\rm{rel}}  \\
&=& 
 \left(-i\frac{G_{\rm F}}{2}\right)^2 \delta_{a'a}\delta_{b'b} i \frac{k_{\rm F}^2}{4\pi^2} 4 \left( 1-\frac{1}{n^2} \right)
 \bar{u}_{q,a'} v\hspace{-0.5em}/ (-4\Lambda-i\pi) \gamma^{0} v\hspace{-0.5em}/ u_{q,a}
   \bar{u}_{P,b'} \frac{1+v\hspace{-0.5em}/}{2} u_{P,b} \nonumber \\
&& + \left(-i\frac{G_{\rm F}}{2}\right)^2 (\vec{\lambda}_{\rm f})_{a'a} \cdot (\vec{\lambda}_{\rm F})_{b'b} \nonumber \\
&& \times i \frac{k_{\rm F}^2}{4\pi^2} \bar{u}_{q,a'} v\hspace{-0.5em}/ \left[ -4\left(n-\frac{4}{n}\right) + n\Lambda^{2} - 2n \log \eta + 2n \log \Lambda + i \pi \frac{4}{n}\right] \gamma^{0} v\hspace{-0.5em}/ u_{q,a} \bar{u}_{P,b'} \frac{1+v\hspace{-0.5em}/}{2} u_{P,b}, \nonumber
\end{eqnarray}
where we find $\log \eta$ as a logarithmically divergent term.
This is again the relativistic version of the Kondo problem for the massless scattering fermion.

\end{appendix}


\end{document}